# Wide-mode-area slow light waveguides in valley photonic crystal heterostructures

CHENGKUN ZHANG,[1,2,*] YASUTOMO OTA,[3] AND SATOSHI IWAMOTO [1,2]

[1]*Research Center for Advanced Science and Technology, The University of Tokyo, 4-6-1 Komaba, Meguro-ku, Tokyo 153-8904, Japan*
[2]*Institute of Industrial Science, The University of Tokyo, 4-6-1 Komaba, Meguro-ku, Tokyo 153-8505, Japan*
[3]*Department of Applied Physics and Physico-Informatics, Faculty of Science and Technology, Keio University, 3-14-1 Hiyoshi, Kohoku-ku, Yokohama-shi, Kanagawa 223-8522, Japan*
*\*zck20@iis.u-tokyo.ac.jp*

**Abstract:** We designed slow-light waveguides with a wide mode area based on slab-type valley photonic crystal (VPhC) heterostructures which are composed of a graphene-like PhC sandwiched by two topologically distinct VPhCs. The group velocity of the topological guided mode hosted in a VPhC heterostructure can be slowed down by shifting the VPhC lattice toward the graphene-like PhC at the domain interfaces. Simultaneously, the mode width of the slow-light topological guided mode can be widened by increasing the size of the graphene-like PhC domain. We found that employing the graphene-like structure at the center domain is crucial for realizing a topological single-guided mode in such heterostructures. Furthermore, the impact of random fluctuations in air-hole size in the graphene-like domain was numerically investigated. Our simulation results demonstrate that the transmittance for the slow-light states can be kept high as far as the size fluctuation is small although it drops faster than that for fast-light states when the disorder level increases. The designed wide-mode-area slow-light waveguides are based on hole-based PhCs, offering novel on-chip applications of topological waveguides.

## 1. Introduction

Topological photonics, a cutting-edge field at the intersection of topology and photonics, explores the creation of topological states of light and novel ways to manipulate light by exploiting them [1-3]. The most famous topological state of light is topological edge states that can be hosted at the interface between two topologically distinct optical structures [4, 5]. The capability of topological edge states to guide light smoothly even under the presence of certain types of structures is attractive for various applications. Various topological edge states in two-dimensional photonic systems, which include optical chiral edge states in systems with broken time-reversal symmetry [6-9], helical edge states in photonic topological insulators [10-14], and valley kink states in valley photonic crystals (VPhCs) [15-19], have been proposed and experimentally demonstrated.

Among them, valley kink states have received much attention since a part of the dispersion curve of the modes is below the light line, allowing light propagation without radiation loss. Additionally, VPhC waveguides exploiting valley kink states can be fabricated using solely dielectric materials like semiconductors. Taking these features, various photonic devices based on VPhC waveguides, including on-chip passive photonic components [20, 21], optical switches [22] and lasers [23, 24], have been demonstrated. Slow-light semiconductor waveguides enabling efficient slow-light guiding even over sharp waveguide turns have been also proposed and demonstrated by exploiting a valley kink state [25, 26]. In general, topological edge states are confined in the region adjacent to the interface between topologically distinct structures, resulting in a limited mode width. Although the tightly localized mode in VPhC-based slow-light waveguides can enhance light-matter interaction, it limits the design flexibility and power transport capacity.

Recently, a PhC heterostructure that can support topological edge states with a large mode width has been proposed [27]. The structure consists of three domains of magnetic PhCs. When two PhC domains at the edges are magnetized oppositely while keeping the center domain unmagnetized, topologically protected chiral edge states with a large mode width emerges within the bandgap of two magnetized PhCs. The topological one-way large-area waveguide can transport electromagnetic field energy efficiently and provide a novel route to control electromagnetic waves. The idea of photonic heterostructures has been extended to VPhC platforms and various devices have been achieved [28-31]. A VPhC heterostructure consisting of a graphene-like PhC sandwiched by two topologically distinct VPhCs allows the construction of a topological triangular-shaped cavity with tunable confinement and high energy transmission capacity, which may be a novel platform for topological lasers [28]. Utilizing tunable mode width of topological guided modes in VPhC heterostructure, a topological concentrator and a topological photonic power splitter have been proposed for photonic integrated circuits [29]. Slow-light waveguides based on VPhC heterostructures are also attractive because of their capability of high-power transmission while keeping slow-light properties. A topological slow light waveguide has been designed in a VPhC heterostructure by modifying the sizes of dielectric rods near the domain walls [32]. However, this topological slow light waveguide is based on rods-in-air PhCs, which is difficult to fabricate and hinders applications in photonic devices.

In this paper, we report a design of a large-mode-width slow-light waveguide based on an air-in-slab VPhC heterostructure, consisting of two topologically distinct hole-based VPhCs and a graphene-like PhC domain sandwiched by the two VPhCs. The mode width of the topological guided mode in the VPhC heterostructure is controlled by changing the width of the graphene-like PhC domain. Besides, by modifying the distance between PhC unit cells forming the domain walls as a tuning knob, the group velocity of the topological guided mode is largely reduced, resulting in the formation of a topological slow-light mode with a large mode width. We also address the importance of the graphene-like structure in the center domain and the impact of random fluctuations in the hole sizes within the graphene-like PhC domain on light propagation properties. These subjects have not been discussed in previous reports on VPhC heterostructures.

## 2. Results and discussion

### 2.1 VPhC heterostructure: structure and topological guided mode

Our designed structure is based on a two-dimensional (2D) VPhC heterostructure $A|B_N|C$, consisting of three domains: a graphene-like PhC B domain with $N$ layers sandwiched by two domains of VPhC A and VPhC C, shown in Fig. 1(a). The domain walls between the graphene-like PhC B and VPhC A, C form zigzag-shape interfaces. Figure 1(b) depicts the unit cells of three domains. Each unit cell is a rhombus-shaped silicon structure ($\varepsilon = 11.56$) with two triangular air holes ($\varepsilon = 1$). The side lengths of large and small air holes in VPhC A and C are $L_L = 1.3a/\sqrt{3}$ and $L_S = 0.7a/\sqrt{3}$, while the side length of all air holes in PhC B is $L_L = L_S = 0.95a/\sqrt{3}$. ($a$ is period of PhC). Figure 1(c) shows TE-mode band structures of three-type unit cells, calculated by 2D plane-wave expansion (PWE) method. Since the inversion symmetry is preserved in PhC B, PhC B hosts a Dirac point at $K(K')$. In VPhC A and C, in contrast, the difference in the side lengths of holes breaks inversion symmetry and lifts the degeneracy at the Dirac point at $K(K')$, resulting in a photonic band gap. The Berry curvature around $K(K')$ in VPhC A and C is non-zero and has an opposite sign of each other, leading to an opposite-signed valley Chern number in each. Consequently, VPhC A and C exhibit different valley phases and can host topological edge modes at their interfaces [15]. The topological edge modes survive after inserting graphene-like PhC B. In addition, thanks to absence of photonic band gap in PhC B, the modes can extend over whole domain of PhC B in VPhC heterostructure [28, 29, 31].

Figure 2(a) shows schematic of VPhC heterostructure A|B$_8$|C, and Fig. 2(b) depicts the corresponding projected band structure. Several guided bands are observed in the bulk band gap. Among them, the topological guided band is marked by a red color, which evolves from the topological edge band in the VPhC zigzag interface, as shown in Fig. 2(d). When $N = 0$, the structure is reduced to a waveguide structure with zigzag interface, which supports only one topological edge state within the band gap (leftmost panel in Fig. 2 (d)). As $N$ increases, the red dispersion continuously changes without gap-closing, which means the topological edge band evolves while keeping its topological origin. Moreover, with increased $N$, more guided bands enter the band gap depicted by black dotted curves, representing nontopological guided modes. Figure 2(c) shows the field distribution of the magnetic field component perpendicular to the PhC plane ($H_z$) for a topological guided mode, indicated by a red point in Fig. 2(b). One can see the field expands over the entire domain of PhC B and decays into domains of VPhC A and C. Thus, a topological guided mode hosted in a VPhC heterostructure can have a wider mode width than that in a conventional VPhC waveguide. The topological guided mode in a VPhC heterostructure is considered to be formed with the topological edge state at the VPhC-based zigzag interface and bulk modes in the graphene-like PhC B, which has been explained based on the $k \cdot p$ perturbation theory [28, 33].

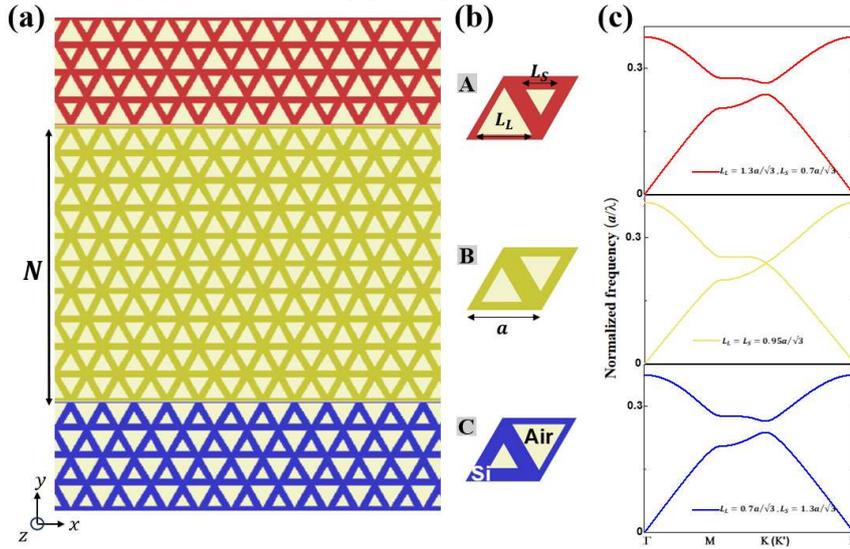

Fig. 1. (a) Schematic diagram of the VPhC heterostructure A|B$_N$|C consisting of three domains, VPhC A, PhC B, and VPhC C. This figure shows a structure with $N$ = 10 as an example. (b) Schematic diagrams for unit cells of VPhC A, PhC B, and VPhC C. (c) Bulk band structures for VPhC A, PhC B, and VPhC C. The calculation is in 2D PWE method.

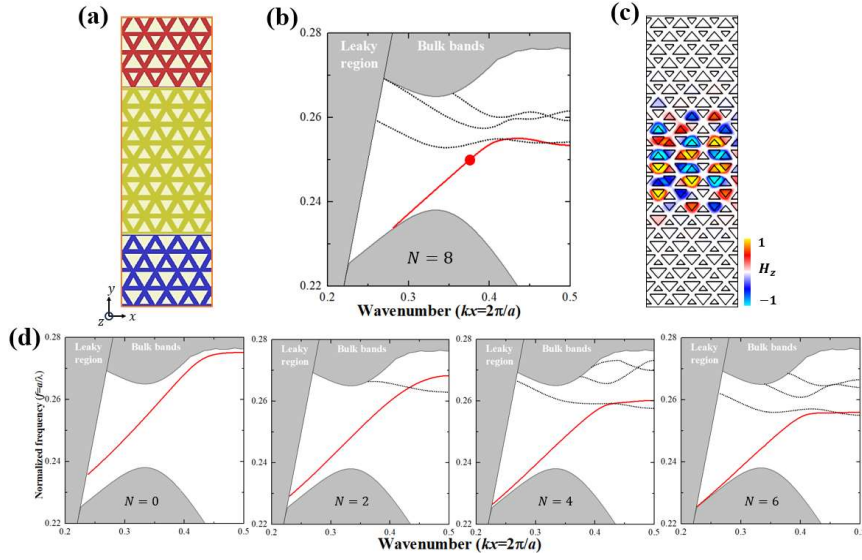

Fig. 2. (a) (b) Schematic diagram of the VPhC heterostructure A|$B_8$|C and corresponding projected band structure. (c) Field distribution of mode marked in (b). (d) Projected band structures of a VPhC zigzag interface ($N = 0$) and VPhC heterostructures A|$B_N$|C with $N = 2,4,6$. The calculation is in 2D PWE method.

*2.2 Slow-light modes in VPhC heterostructures*

To achieve slow-light modes in VPhC heterostructures, we modify the distance $d$ between two rows of air holes near the boundary between graphene-like PhC B and VPhCs. The schematic diagrams of VPhC heterostructure A|$B_8$|C with $d = 1.4a$ (unmodified) and $d = 1.15a$ (modified) are depicted in Fig. 3(a). Despite the connection of two rows of air holes in the modified VPhC heterostructure, the semiconductor rods remain interconnected, ensuring the structure can still be fabricated. Figure 3(d) shows the projected band structures for VPhC heterostructures ($N = 8$) with different $d$ values. As the distance $d$ reduces from $1.4a$ to $1.15a$, the topological guided mode, depicted by red curves, undergoes continuous evolution without any gap closure or alteration in its topological characteristic. Interestingly, the topological guided mode tends to have smaller group velocities and becomes a slow-light mode near the K point ($kx = 0.66 \pi/a$) as $d$ decreases. Figure 3(b) shows the field distribution of the $H_z$ component of the topological slow-light mode marked as Mode I by a blue point in Fig. 3(d). The field distributes over the entire domain of graphene-like PhC B and decays in domains of VPhCs A and C, which indicates Mode I is a topological slow-light guided mode with a large mode width. The appearance of a slow-light region in the topological guided mode can be understood as follows. The reduction of $d$ makes the effective refractive index near the domain walls smaller, shifting the topological guided mode to higher frequencies in most of the Brillouin zone. On the other hand, the mode frequency at the Brillouin zone edge is less affected by the local structural modulation since the mode field at the Brillouin zone edge (Mode II marked with a green circle in the rightmost panel of Fig. 3 (d)) localizes more tightly in graphene-like PhC B than the modes apart from the Brillouin zone edge as shown in Fig. 3 (c). Consequently, the topological band becomes flatter.

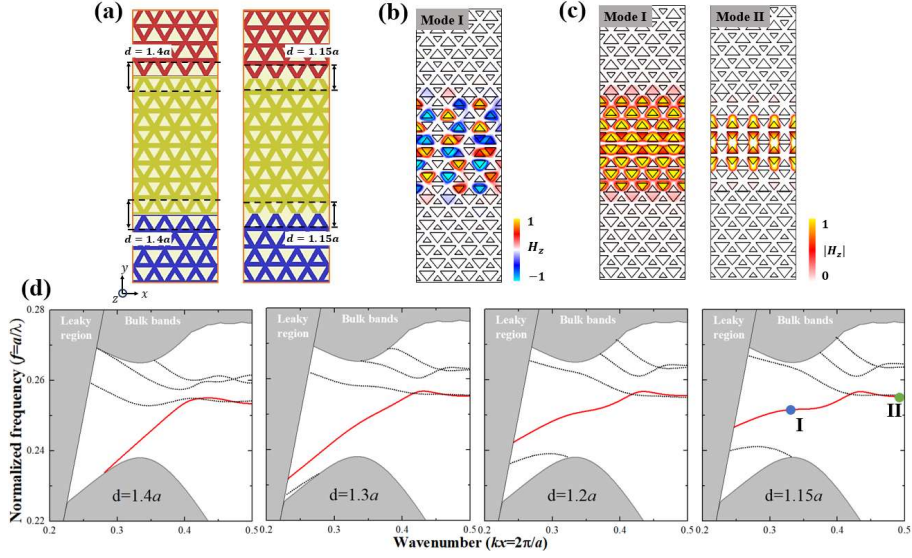

Fig. 3. (a) Schematic diagrams of the VPhC heterostructures A|B$_8$|C with different distance values $d = 1.4a$ and $d = 1.15a$. (b) Field distribution of Mode I in (d). (c) Amplitude distribution of Mode I and II in (d). (d) Projected band structures of VPhC heterostructures A|B$_8$|C with different $d$ values. The calculation is in 2D PWE method.

Next, we consider VPhC heterostructures with the same distance value $d = 1.15a$ but different $N$ values depicted in Fig. 4 (a). Figure 4(c) shows the corresponding projected band structures, at which red dispersions and black dotted dispersions represent topological guided modes and nontopological guided modes, respectively. The slow-light region in the topological guided band exists in VPhC heterostructures with different $N$, indicating that the control of group velocity by modifying $d$ value works for VPhC heterostructure even as $N$ changes. It is also worth noting that the slow light is single mode except for the case of $N$ = 4. Black balls in Fig. 4 (c) mark the slow-light mode whose group index $N_g$ is about 50 in different structures. The corresponding amplitude profiles of the $H_z$ component are shown in Fig. 4(b). These modes spread over the domain of PhC B and the mode width becomes wider as $N$ increases, which indicates that the mode width of the topological slow-light mode hosted in VPhC heterostructures can be tuned by changing $N$.

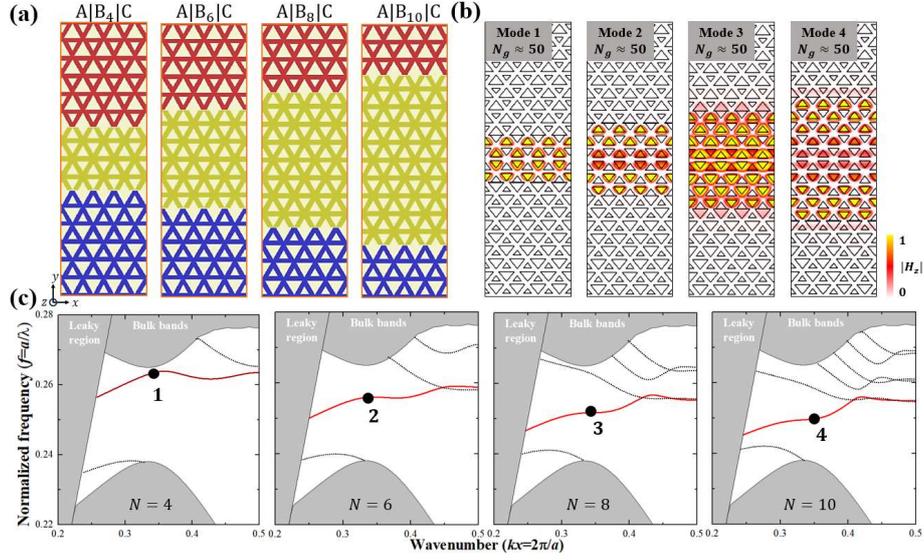

Fig. 4. (a) Schematic diagrams of VPhC heterostructure $A|B_N|C$ with different PhC B layers ($N = 4, 6, 8, 10$) and same distance value $d = 1.15a$. (b) Amplitude distribution of modes marked in (c) with similar group index ($n_g \approx 50$). (c) Projected band structures of VPhC heterostructure $A|B_N|C$ with same distance value $d = 1.15a$ and different PhC B layers ($N = 4, 6, 8, 10$). The calculation is in 2D PWE method.

## *2.3 Significance of graphene-like structure in the domain B*

In this section, we discuss the importance of graphene-like structure in VPhC heterostructures by replacing the graphene-like PhC B with a uniform layer. Figure 5(a) shows the schematic diagram of VPhC A/uniform layer/VPhC C heterostructure we investigated, of which the graphene-like PhC B domain is replaced with an artificial uniform layer of refractive index $n = 2.9$ compared to $A|B_8|C$. For a more equitable comparison between two different heterostructures, we have aligned the refractive index of the uniform layer with the average value in the PhC B domain. We plot the projected band structure of the A/uniform layer/C heterostructure in Fig. 5(b). The band structure is largely different from the band structure for the VPhC heterostructure with $N = 8$ (rightmost panel in Fig. 3 (d)). In the heterostructure with the uniform layer as core region, there are many guided modes within the bulk band gap. Some modes have large group indices and a wide mode width (see Fig. 5 (c) for example), but they are not isolated from other modes in frequency, which is disadvantageous in some applications compared to single-mode slow light obtained in VPhC heterostructures with a graphene-like domain. This result indicates that the use of a graphene-like structure in Domain B is crucial to have a single-mode slow-light mode with a wide mode width in VPhC heterostructures.

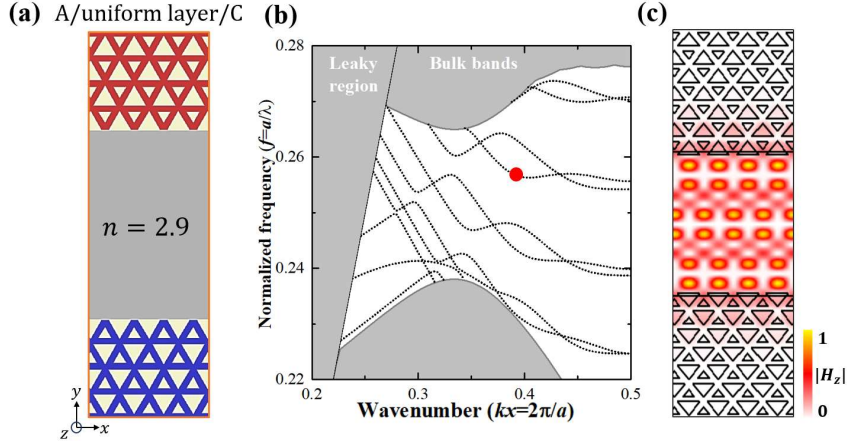

Fig. 5. (a) Schematic diagram of heterostructure VPhC A/uniform layer/VPhC C. (b) Projected band structure in A/uniform layer/C heterostructure. (c) Amplitude distribution of mode marked as red point in (b). The calculation is in 2D PWE method.

*2.4 Robustness of topological guided modes in VPhC heterostructures*

Generally, robustness of topological guided modes in VPhC waveguides and VPhC heterostructures is demonstrated by showing smooth propagation against some large disorder like bending [17, 26, 29, 32, 34]. However, for practical use of topological waveguides, it is important to understand the robustness against random fluctuations in structures from the designed ones, which are unavoidable in real devices [35-38]. In this section, we discuss the robustness of the topological guided mode in a VPhC heterostructure against such random structural disorders by examining light propagation through a VPhC heterostructure with random structural fluctuations.

We computed light transmittance in a VPhC heterostructure $A|B_8|C$ based waveguide ($d = 1.15a$) with a random size fluctuation in some holes. The schematic of the waveguide structure investigated here is depicted in Fig. 6(a), where some triangular air holes in the graphene-like PhC B domain are replaced by triangular air holes with random side lengths (brown triangles), forming a disorder region. The side length $L$ of each brown triangle was randomly determined by following Normal distribution with average value $\mu$ and standard deviation $\sigma$, $L = rand[Norm(\mu, \sigma^2)]$. We fixed $\mu = 0.95(a/\sqrt{3})$ the same as the side length of air holes in the unperturbed graphene-like PhC B unit cell and tuned $\sigma$ to control disturbance level. In the simulation, we focus on the frequency range of the topological single mode to avoid the influence of the interplay between multiple modes. Based on the frequency dependence of the group index of the topological single-guided mode, computed by 2D finite-difference time-domain method (2D-FDTD) and illustrated in Fig. 6 (b), we categorize the mode into the fast-light region ($n_g < 15$) and the slow-light region ($n_g \geq 15$) for further analysis.

Figure 7(a) plots transmission spectra of VPhC heterostructure waveguides with disturbance $\sigma = 0.01, 0.05, 0.1$ (black curves) and without disturbance (red curve). The fast-light and slow-light regions are shaded by blue and orange colors, respectively. The transmission spectra are calculated by 2D-FDTD method. As seen from the red curves in Fig. 7(a), the topological single modes in VPhC heterostructure waveguides without the size fluctuation region exhibit almost perfect transmittance over both the fast- and slow-light regions. For $\sigma = 0.01$, the fast-light region and most of the slow-light region keep high transmittance, showing robustness against the disturbance with this level. A dip in the transmission spectrum is observed near $f = 0.2505$ close to the frequency showing the highest group index in Fig. 6(b). This suggests that the mode with extremely low group velocity is sensitive to disturbance in the waveguides. For $\sigma = 0.05$, the transmission in the fast-light region and some portions of the slow-light region are slightly reduced but still rather high. On the other hand, the dip around $f = 0.2505$ becomes wider and deeper. When σ increases to 0.1, the transmittance in the slow-light region drops dramatically while high transmission is kept at some frequencies in the fast-light region. We select a fast-light mode ($f = 0.245$) and a slow-light mode ($f = 0.25$) for comparison and plot the steady state field distribution at each frequency in Fig. 7(b) and (c) respectively. The boxes with a dashed white line indicate the disordered region composed of air holes with random size fluctuation. It is observed that the fast-light mode can smoothly propagate through the disordered region without significant backscattering showing strong robustness, while the slow-light mode is blocked at high disturbance $\sigma = 0.1$ showing weaker robustness.

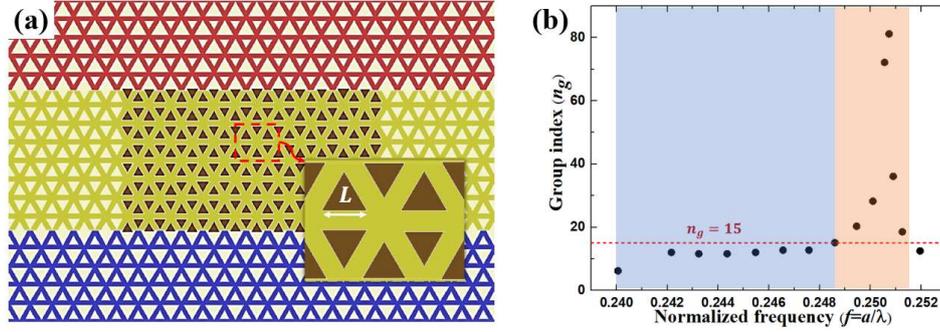

Fig. 6. (a) Schematic diagram of VPhC heterostructure $A|B_8|C$ based waveguide ($d = 1.15a$) and partial air holes are set with random side lengths marked by brown color. (b) Group index of the topological single mode in VPhC heterostructure $A|B_8|C$ with $d = 1.15a$. The fast-light and slow-light regions are shaded by blue and orange colors. The calculation is in 2D FDTD method.

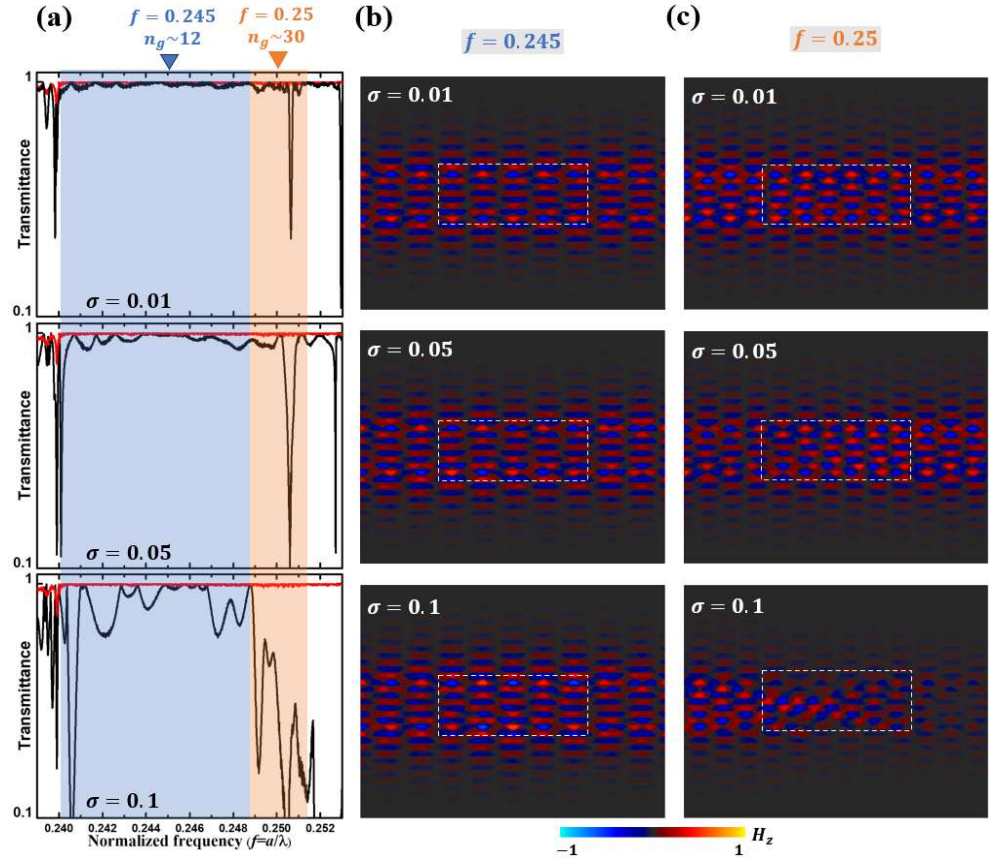

Fig. 7. (a) Transmittance of VPhC heterostructure $A|B_8|C$ based waveguide ($d = 1.15a$) with disturbance $\sigma = 0.01, 0.05, 0.1$. The fast-light and slow-light regions are shaded by blue and orange colors. (b) (c) Steady-state field distributions for fast-light mode ($f = 0.245$) and slow-light mode ($f = 0.25$) in VPhC heterostructure with disturbance $\sigma = 0.01, 0.05, 0.1$. The calculation is in 2D FDTD method.

To assess the impact of random disorders more quantitatively, in Fig. 8, we plotted the average transmittance at the normalized frequencies $f$ of 0.245 and 0.25, which belong to the fast- and slow-light regions, respectively. The group index is ~12 and ~30 at each frequency. For each $\sigma$, the average values were computed from the results for five different configurations. At low disturbance levels ($\sigma < 0.05$), the average transmittances are kept in high values in both frequencies, demonstrating robustness against random size fluctuations. Considering that VPhC waveguides with a small hole size fluctuation ($\sigma < 0.01$) have been fabricated in a controlled manner in [38], it is likely that the state-of-the-art fabrication technology can realize the VPhC heterostructure waveguides with a size fluctuation with $\sigma < 0.05$. As the disturbance increases, the average transmittance at both regions drops gradually. Notably, the average transmittance at $f = 0.25$ in the slow-light region drops quicker than that at $f = 0.245$ in the fast-light region. The average transmittance becomes less than 0.6 when $\sigma > 0.1$ at $f = 0.25$ and $\sigma > 0.12$ at $f = 0.245$. Our numerical results agree with the recently reported transmittance measurement for VPhC interface waveguides with random size fluctuations [38]. The results show the topological fast-light modes have higher backscattering mean free path (BMFP) than topological slow-light modes, indicating a group index dependence of robustness [38]. The different behavior of fast-light and slow-light modes to random size fluctuations can be attributed to the difference in group velocity. When light propagates in a disordered region, slow-light modes will suffer from scattering more, compared to fast-light modes due to low group velocity.

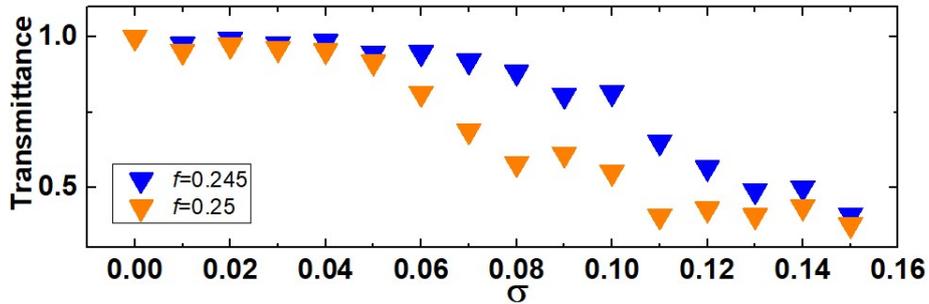

Fig. 8. Average transmittance as a function of disturbance $\sigma$ for fast-light mode ($f = 0.245$, blue) and for slow-light mode ($f = 0.25$, orange). The calculation is in 2D FDTD method.

## 3. Conclusion

We proposed a design of VPhC heterostructure supporting a topological slow-light mode with a large mode width. In detail, the VPhC heterostructure is an air-in-slab structure, realized by a graphene-like PhC domain sandwiched by two topologically distinct VPhC domains, supporting a topological guided mode with a tunable mode width depending on the size of the graphene-like PhC domain. Shrinking the distance between the unit cells forming the domain boundaries slows down the group velocity of the topological guided mode, resulting in single-mode slow-light states with a broad mode width. We uncovered the importance of the graphene-like structure to have such single-mode topological states in VPhC heterostructures by comparing the projected band structures for a VPhC heterostructure with a graphene-like layer, and a uniform layer sandwiched with two VPhCs. Finally, we numerically investigated the impact of the random fluctuation in the size of air holes in the graphene-like PhC domain on light transmission properties. The results indicate the topological guided mode is reasonably robust against a disturbance with a standard deviation of less than 0.05 both for the fast- and slow-light modes. Our work will provide a new design concept for various kinds of slow-light devices including slow-light waveguides for high-power transmission, broad-area single-mode lasers, and highly sensitive optical sensors.

**Funding.** This work is supported by JST CREST(JPMJCR19T1), KAKENHI (22H00298, 22H01994) and Asahi Glass Foundation.

**Acknowledgments.** The authors thank Mr. Lu Guangtai for useful discussions.

**Disclosures.** The authors declare no conflicts of interest.

**Data availability.** Data underlying the results presented in this paper are not publicly available at this time but may be obtained from the authors upon reasonable request.